\documentclass[10pt,conference]{IEEEtran}
\IEEEoverridecommandlockouts

\usepackage{cite}
\usepackage{amsmath,amssymb,amsfonts}
\usepackage{algorithmic}
\usepackage{graphicx}
\usepackage{textcomp}
\def\BibTeX{{\rm B\kern-.05em{\sc i\kern-.025em b}\kern-.08em
    T\kern-.1667em\lower.7ex\hbox{E}\kern-.125emX}}

\usepackage{packages}

\begin{document}

\title{From Expectation to Habit: Why Do Software Practitioners Adopt Fairness Toolkits?

}

\author{\IEEEauthorblockN{Gianmario Voria}
\IEEEauthorblockA{University of Salerno\\
Salerno, Italy\\
gvoria@unisa.it}
\and
\IEEEauthorblockN{Stefano Lambiase}
\IEEEauthorblockA{University of Salerno\\
Salerno, Italy\\
slambiase@unisa.it}
\and
\IEEEauthorblockN{Maria Concetta Schiavone}
\IEEEauthorblockA{University of Salerno\\
Salerno, Italy\\
m.schiavone29@studenti.unisa.it}
\and
\IEEEauthorblockN{Gemma Catolino}
\IEEEauthorblockA{University of Salerno\\
Salerno, Italy\\
gcatolino@unisa.it}
\and
\IEEEauthorblockN{Fabio Palomba}
\IEEEauthorblockA{University of Salerno\\
Salerno, Italy\\
fpalomba@unisa.it}}

\maketitle

\begin{abstract}
As the adoption of machine learning (ML) systems continues to grow across industries, concerns about fairness and bias in these systems have taken center stage. Fairness toolkits—designed to mitigate bias in ML models—serve as critical tools for addressing these ethical concerns. However, their adoption in the context of software development remains underexplored, especially regarding the cognitive and behavioral factors driving their usage. As a deeper understanding of these factors could be pivotal in refining tool designs and promoting broader adoption, this study investigates the factors influencing the adoption of fairness toolkits from an individual perspective. Guided by the Unified Theory of Acceptance and Use of Technology (UTAUT2), we examined the factors shaping the intention to adopt and actual use of fairness toolkits. Specifically, we employed Partial Least Squares Structural Equation Modeling (PLS-SEM) to analyze data from a survey study involving practitioners in the software industry. Our findings reveal that performance expectancy and habit are the primary drivers of fairness toolkit adoption. These insights suggest that by emphasizing the effectiveness of these tools in mitigating bias and fostering habitual use, organizations can encourage wider adoption. Practical recommendations include improving toolkit usability, integrating bias mitigation processes into routine development workflows, and providing ongoing support to ensure professionals see clear benefits from regular use.
\end{abstract}


\section{Introduction}

Machine Learning (ML) has become pervasive, with its adoption accelerating across a wide array of industries and everyday applications \cite{zhou2018human}. ML-enabled systems—software systems powered by AI or ML algorithms \cite{ai_enabled}—are revolutionizing sectors such as healthcare and entertainment by improving efficiency, optimizing decision-making processes, and driving innovative solutions \cite{wang2021comparative,ni2020survey,miller2015can,olson2011algorithm}.

As ML continues to spread, it has also prompted significant ethical concerns, particularly around \emph{fairness} \cite{mehrabi2021survey}, which refers to the principle that models should make impartial decisions, avoiding bias or discrimination against certain groups. Unfairness occurs when models inherit biases present in training data \cite{pagano2023bias,pessach2022review}, resulting in decisions that undermine trust and pose ethical as well as legal challenges \cite{miller2019machine}. This is proven by several known ethical incidents caused by ML applications, e.g., Facebook vision model that put the ``primate" label to black men or Amazon assigning lower sales ranking to books containing LGBTQIA+ themes, highlighting the urgent need for fair ML software \cite{brun2018software,facebookprimates,amazongay,ia_ethical_incidents}.

Acknowledging the critical importance of fairness, the software engineering (SE) research community—particularly in the domain of software engineering for artificial intelligence (SE4AI)—has made substantial strides in developing bias mitigation techniques \cite{hort2024bias}, recognizing fairness as a crucial non-functional requirement. These approaches can generally be grouped into three main categories: \emph{pre-processing}, \emph{in-processing}, and \emph{post-processing} techniques. In this regard, the research community and organizations have developed instruments to make these solutions available for \emph{software practitioners}, e.g., \textsc{Aif360} \cite{aif360} or \textsc{FairLearn} \cite{fairlearn}. These tools referred to as \emph{fairness toolkits}, comprise ready-to-use metrics to measure fairness or bias mitigation techniques \cite{toolkit_survey}.

While fairness toolkits have proven effective in mitigating bias~\cite{toolkit_survey,toolkit_landscape}, there is still a significant gap in understanding their actual adoption. Specifically, it remains unclear what \textit{decision-making heuristics} lead practitioners to consider using fairness toolkits in their workflow. We argue that this is an important limitation for two reasons. First, studying the adoption of fairness toolkits may uncover the main drivers that need to be considered or encouraged to further increase the uptake of these tools, as suggested by previous research investigating technology acceptance \cite{lambiase2024investigating, tamilmani2020utautreview}. Second, understanding the considerations that lead to their adoption can offer additional insights into how existing fairness toolkits can be refined and better integrated into practitioners' workflows~\cite{organizational_responsible_ai}. This may inform recommendations for designing the next generation of fairness toolkits. In summary, a deeper understanding of these heuristics could provide valuable insights for both researchers and toolkits vendors.

Recognizing the aforementioned opportunities, our goal is to offer a complementary perspective by investigating the key factors influencing practitioners' willingness to adopt fairness toolkits. Therefore, this research seeks to address this gap, starting by defining the following guiding research question:

\steResearchQuestionBox{\faBullseye \hspace{0.05cm} \textbf{Research Question.} \textit{What factors influence software practitioners in the adoption of fairness toolkits?}}

To address the research question, we conducted a quantitative study grounded in the Unified Theory of Acceptance and Use of Technology (UTAUT2) \cite{utaut2}. We surveyed expert practitioners and employed Partial Least Squares Structural Equation Modeling (PLS-SEM) for data analysis \cite{hair_2014_PLS}. Our results show that software practitioners' intention to adopt fairness toolkits is mainly driven by their expectancy of the performance of these instruments, i.e., the extent to which they are able to mitigate bias. Moreover, habit emerged as a driver for both the intention to use and the actual adoption of fairness toolkits by practitioners.

\section{Background and Related Works}
In the following subsections, we summarize the most relevant literature regarding fairness.

\textbf{Machine Learning Fairness.} Fairness in decision-making refers to the absence of bias or favoritism based on inherent or acquired characteristics~\cite{pessach2022review, mehrabi2021survey, starke2021fairness}. Various metrics and strategies evaluate fairness in ML, focusing on data similarities, decision probabilities, and cause-effect relationships~\cite{verma2018fairness}. Majumder et al.~\cite{fair_enough} categorized fairness metrics into seven groups, although not all nuances are captured.

Bias in ML systems can be mitigated using pre-, in-, and post-processing techniques. Pre-processing targets bias in training data, with Sharma et al.\cite{sharma2020data} and Calmon et al.\cite{calmon2017optimized} using probabilistic methods, and Chakraborty et al.\cite{chakraborty2021bias} introducing \textsc{Fair-SMOTE}. In-processing adjusts the learning algorithm itself; Zhang et al.\cite{zhang2018mitigating} used adversarial methods, and Kamishima et al.\cite{kamishima2012fairness} applied regularization. Reweighting methods, like those from Kamiran and Calders\cite{kamiran2012data} and Chakraborty et al.\cite{chakraborty2020fairway}, adjust instance weights. Post-processing techniques refine model outputs after training, with Galhotra et al.\cite{galhotra2017fairness} introducing \textsc{Themis} and Udeshi et al.~\cite{udeshi2018automated} developing \textsc{Aequitas}.

\textbf{Software Engineering for ML Fairness.} Fairness in ML has gained traction in the SE community, with various studies addressing it from multiple perspectives~\cite{pessach2022review, mehrabi2021survey, starke2021fairness, chen2024fairness, hort2024bias}. Ferrara et al.\cite{ferrara2024refair} stressed the importance of context-aware fairness requirements. Additionally, Ferrara et al.\cite{ferrara2024fairness} advocated for fairness integration across the development lifecycle. Discrimination often stems from biased training datasets~\cite{vasudevan2020lift}. Zhang and Harman~\cite{zhang2021ignorance} argued that increasing dataset features does not inherently reduce discrimination, while Chakraborty et al.\cite{chakraborty2021bias} emphasized the role of feature selection. Sesari et al.\cite{sesari2024understanding} highlighted the need to examine fairness across the entire dataset, and Voria et al.~\cite{voria2024mapping, voria2024survey} cataloged fairness-aware practices surveying domain experts.

\textbf{Fairness Toolkits in Practice.} Various open-source fairness toolkits help researchers and developers create fairer ML models~\cite{fairnessToolkitCheckbox}. \textsc{AIF360}\cite{aif360}, developed by IBM, offers a comprehensive suite of fairness metrics and bias mitigation techniques. \textsc{Fairlearn}\cite{fairlearn} is a Python-based library focused on fairness assessment and mitigation using in-processing techniques. Google’s \textsc{What-If Tool}\cite{whatif} emphasizes fairness and explainability through interactive analysis, while \textsc{Scikit-fairness}\cite{skfairness} extends scikit-learn with bias analysis tools.

Lee and Singh~\cite{toolkit_landscape} examined the misalignment between current open-source fairness toolkits and practitioners' needs through focus groups, interviews, and surveys, identifying gaps that necessitate improved support for implementing fairness. Similarly, Holstein et al.\cite{holstein_industry} documented challenges in developing fair ML systems within commercial teams based on interviews and surveys. Deng et al.\cite{toolkit_survey} conducted an empirical investigation into industry practitioners' engagement with fairness toolkits, identifying usability and effectiveness improvements through think-aloud interviews and surveys. Abstracting from the technical perspective, Rakova et al.~\cite{organizational_responsible_ai} explored fairness issues organizationally, developing a framework to analyze how organizational culture and structure impact responsible software initiatives. They identified challenges and enablers through interviews, mapping current structures to ideal future processes.

\smallskip
\stesummarybox{\faList \hspace{0.05cm} Our Contribution.}{The objective of our work aligns with existing studies, specifically focusing on the practical use and effectiveness of fairness toolkits. However, it distinguishes itself by incorporating the UTAUT (Unified Theory of Acceptance and Use of Technology) model as a framework to systematically understand how these tools are utilized and perceived by users. This integration allows for an examination of user adoption and acceptance of technology, along with the identification of factors that influence the effectiveness and usability of fairness toolkits, thus filling a crucial gap in the current literature.}

\begin{figure}
    \centering
    \includegraphics[width=1\linewidth]{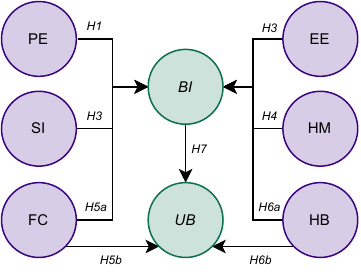}
    \caption{Overview of the UTAUT2 theoretical model with our hypothesis.}
    \label{fig:utautmodel}
\end{figure}

\section{Hypothesis and Theory Development}
\label{sec_theory}

This study seeks to identify the key individual factors that influence the adoption of fairness toolkits in software development and engineering. Given the lack of studies that specifically investigate the factors influencing the adoption of fairness toolkits, we have chosen to base our approach on the \textsc{Unified Theory of Acceptance and Use of Technology (UTAUT2)}~\cite{utaut2} model. This theory is widely regarded as one of the most comprehensive frameworks for examining technology adoption across diverse contexts \cite{tamilmani2020utautreview,venkatesh2000utautgender,lambiase2024investigating,utaut2}, enabling us to minimize bias and build on established knowledge. 

\subsection{The Unified Theory of Acceptance and Use of Technology}

One of the theoretical models developed to predict technology adoption and use is \textsc{UTAUT} \cite{utaut}. This model posits that the actual usage \textbf{(UB)} of technology is determined by behavioral intention \textbf{(BI)}. The perceived likelihood of adopting the technology is influenced by the direct effects of four key constructs: the belief that using the system will enhance job performance (Performance Expectancy, \textbf{PE}), the perceived ease of using the system (Effort Expectancy, \textbf{EE}), the perception that organizational and technical infrastructures are in place to support system use (Facilitating Conditions, \textbf{FC}), and the perception that important others believe the system should be used (Social Influence, \textbf{SI}). Individual-related factors, such as age, gender, and experience, are typically considered to moderate or diminish the relationship between technology use and behavioral intention\cite{utautnew}.

Despite the widespread acceptance of \textsc{UTAUT}, Venkatesh et al.~\cite{utaut2} later introduced \textsc{UTAUT2}, an updated version of the original model that includes three additional constructs, emphasizing the user as a customer and stakeholder rather than merely a technology adopter~\cite{utaut2}. This extension was designed to provide greater precision in explaining user behavior. \textsc{UTAUT2} provides the following additional constructs: the degree of pleasure or enjoyment derived from using the technology (Hedonic Motivation, \textbf{HM}), the cognitive trade-off between the perceived benefits of the technology and its monetary cost (Price Value, \textbf{PV}), and the extent to which individuals tend to perform behaviors automatically through learning (Habit, \textbf{HB}).

\textbf{Motivation and Choices.}
Given our objective to investigate the adoption of fairness toolkits by software practitioners, the \textsc{UTAUT2} model was a natural selection. In comparison to its predecessors, the Technology Acceptance Model (\textsc{TAM})~\cite{davis1989technology} and \textsc{UTAUT}~\cite{utaut}, the new model encompasses a broader range of individual-level factors that capture various dimensions of technology adoption~\cite{utaut2}. Moreover, the constructs of social influence and facilitating conditions allow us to consider environmental factors that may affect the adoption process~\cite{utaut2}. Importantly, \textsc{UTAUT2} provides us with established and validated measurement instruments, enabling us to contextualize our findings within a substantial body of literature utilizing the same theoretical framework.

The standard \textsc{UTAUT2} model includes three moderating variables: age, gender, and experience. However, in the interest of model parsimony, previous research~\cite{venkatesh2000utautgender, lambiase2024investigating} that adopted the \textsc{UTAUT} framework in software engineering research chose to exclude these three. To ensure reliability and robustness, we still conducted a preliminary analysis using the moderating variables. The results, available in our online appendix~\cite{appendix}, revealed that these three were not significant. Hence, according to previous research, we made the same decision to exclude the moderators from our final analysis.

Finally, the construct Price Value (PV)\cite{utaut2} has been excluded. It evaluates the perception of the relationship between the price paid and the benefits gained from using the technology. However, since fairness toolkits are mostly open-source and thus freely accessible without any direct economic cost to the user, this construct was deemed insignificant.

\subsection{Hypotheteses Development}
In the following, we present the hypothesis we developed for the constructs of the \textsc{UTAUT2} model to understand software practitioners' intention to use and actual usage of fairness toolkits. Figure \ref{fig:utautmodel} summarizes the model built upon the hypotheses described in this section.

One of the constructs of the model is Performance Expectancy, which refers to the degree to which an individual believes that adopting a particular technology will enhance their job performance~\cite{utaut}. In essence, this construct suggests that practitioners are more likely to utilize fairness toolkits if they perceive these tools as beneficial for completing their routine software development tasks, which in this case may be influenced by how much they are expected to produce fair software for their organization\cite{organizational_responsible_ai}. Positive outcomes associated with fairness toolkits—such as improved efficiency in identifying biases, enhanced accuracy in decision-making, and strengthened capabilities for addressing ethical dilemmas—can significantly influence practitioners' willingness to integrate these tools into their workflows\cite{figueroa2022ethicalperformanceexpectancy}. Given the potential of fairness toolkits to streamline development processes and provide substantial performance benefits, it is reasonable to propose that performance expectancy plays a role in practitioners' intentions to adopt these technologies~\cite{oneto2020mlfairnessbook, figueroa2022ethicalperformanceexpectancy}. Therefore, we hypothesize that \textit{\textbf{(H1: PE$\rightarrow$BI)} Performance Expectancy (PE) positively influences the intention to adopt (BI) fairness toolkits by software practitioners.}

The degree of ease with which a particular technology can be utilized is referred to as Effort Expectancy \cite{utaut}. Individuals are more likely to adopt new technologies when they find them easy to understand and use\cite{davis1989perceived}. Stakeholders may decide whether to incorporate fairness toolkits into their development processes based on how straightforward it is to integrate and utilize these tools\cite{toolkit_survey}. We anticipate that effort expectancy will positively influence the intention to adopt fairness toolkits for software development, given the significance of ease of use and reduced cognitive load in technology acceptance. Therefore, we hypothesize that \textit{(\textbf{H2: EE$\rightarrow$BI)} Effort Expectancy (EE) positively influences the intention to adopt (BI) fairness toolkits by software practitioners.}


Social Influence refers to the degree to which an individual perceives that important others believe they should use a new system\cite{utaut}. According to this concept, individuals are more likely to adopt new technologies if they feel that their colleagues, supervisors, or social norms advocate for their use. Stakeholders may receive approval and encouragement from peers, mentors, or industry leaders, which can significantly impact their decision to adopt fairness toolkits~\cite{figueroa2022ethicalperformanceexpectancy}, given the critical impact that fairness may have on society. Therefore, we hypothesize that \textit{\textbf{(H3: SI$\rightarrow$BI)} Social Influence (SI) positively influences the intention to adopt (BI) fairness toolkits by software practitioners.}


Hedonic Motivation refers to the pleasure or enjoyment derived from using fairness toolkits for development tasks\cite{utaut2}. This construct significantly influences user acceptance of technology. It can be observed that the more enjoyable an activity is, the more positive the practitioner’s attitude toward it becomes~\cite{van2004user}. Therefore, we hypothesize \textit{\textbf{(H4: HM$\rightarrow$BI)} Hedonic Motivation (HM) positively influences the intention to adopt (BI) fairness toolkits by software practitioners.}


The belief of an organizational and technical infrastructure that supports the use of a particular technology is known as Facilitating Conditions \cite{utaut}. Practitioners are more likely to adopt fairness toolkits if they have access to the necessary tools, as well as support and training. They are also more inclined to actually utilize them if they are confident that their organization provides the resources to integrate and use them effectively. Therefore, we hypothesize that \textit{Facilitating Conditions (FC) \textbf{(H5a: FC$\rightarrow$BI)} positively influences the intention to adopt (BI) fairness toolkits by software practitioners and \textbf{(H5b: FC$\rightarrow$UB)} positively influences the actual use behavior (UB) regarding fairness toolkits by practitioners.}

Habit refers to the extent to which individuals tend to act automatically as a result of learning\cite{utaut2}. Practitioners are more likely to adopt fairness toolkits if they are familiar with certain tools and use them regularly. Furthermore, habit influences actual usage behavior, as practitioners who consistently integrate new tools are more inclined to use fairness toolkits consistently in their development activities. Therefore, we hypothesize that \textit{Habit (HB) \textbf{(H6a: HB$\rightarrow$BI)} positively influences the intention to adopt (BI) fairness toolkits by software practitioners and \textbf{(H6b: HB$\rightarrow$UB)} positively influences the actual use behavior (UB) regarding fairness toolkits by software practitioners.}

Lastly, according to psychological models, individual behavior can be predicted and conditioned by personal intentions. Hence, UTAUT2 posits that behavioral intention significantly impacts actual technology usage \cite{utaut}. Based on this premise, we hypothesize that \textit{\textbf{(H7: BI$\rightarrow$UB)} Behavioral Intention (BI) to use fairness toolkit positively influences the actual use behavior (UB) of software practitioners.}


We combined the hypotheses outlined above with those from the UTAUT2 framework to explore software practitioners' adoption of fairness toolkits, as shown in Figure \ref{fig:utautmodel}.

\section{Research Design}

To evaluate the above-mentioned hypotheses and explore software practitioners' intentions to adopt fairness toolkits, we conducted a survey study. 

\begin{figure}
\centering
\includegraphics[width=1\linewidth]{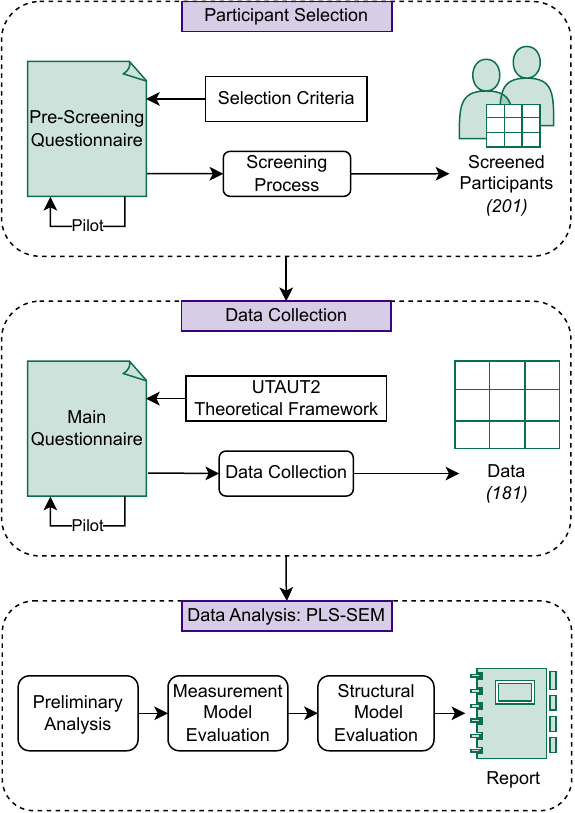}
\caption{Research Method.}
\label{fig:method}
\end{figure}

Figure \ref{fig:method} presents an overview of our research methodology, which we elaborate upon in this section. We initiated our process by carefully defining participant selection criteria for our survey and calculating the sample size using G*Power~\cite{faul_2009_GPower}, taking into account the complexities of our theoretical model. Subsequently, we developed questionnaires grounded in validated instruments from the literature~\cite{utaut2} to measure the model's constructs. Throughout the questionnaire development process, we adhered strictly to established guidelines, incorporating iterative pilot testing to ensure the highest standards of validity and reliability~\cite{kitchenham2008_PersonalOpinionSurveys, andrews2007_survey_guidelines}.

Participants first completed a carefully designed screening questionnaire, followed by a second survey intended to assess the \textsc{UTAUT2} constructs. We then rigorously analyzed the collected data using \textit{Partial Least Squares-Structural Equation Modeling (PLS-SEM)}~\cite{hair_2014_PLS} to address our research questions. PLS-SEM is a versatile statistical method that integrates factor analysis and multiple regression to analyze complex relationships between constructs~\cite{hair_2014_PLS, russo2021_pls_SLR}. It is especially useful for predictive analysis, theory development, and models with multiple independent and dependent variables, particularly when dealing with small sample sizes or non-normal data distributions.

\subsection{Participant Selection and Demographics}
To gather data for this study, we implemented a survey utilizing a cluster sampling strategy via Prolific, a reputable academic data collection platform.\footnote{Prolific (\url{www.prolific.com}) [October 2024]} Prolific's advanced filtering capabilities allowed us to precisely define specific criteria for selecting potential participants that aligned with our research requirements.

We formulated an ideal participant profile based on our study's objectives, focusing on individuals demonstrating proficiency in computer programming and actively engaged in the information technology industry. Leveraging Prolific's filters, we targeted participants meeting these exacting criteria. The survey, titled \textit{[``Fairness in Software Development] Pre-Screening — Fairness Toolkit Adoption"}, clearly stated its goals in the description, ensuring that only qualified respondents, including those not currently employed as developers but meeting the pre-screening standards, participated.

To further enhance the reliability and accuracy of our data, we implemented two additional filters. First, we required that participants be fluent in English, as the survey was conducted exclusively in this language. Second, we required participants to have a 100\% approval rate from previous surveys. On Prolific, researchers can reject responses based on specific criteria, and a participant's approval rate reflects the percentage of submissions that have been accepted, serving as a reliable proxy for their consistency in providing high-quality data.
In addition to these filters, we incorporated questions developed by Danilova et al.~\cite{danilova2021_developers_questions} to assess participants' programming knowledge, ensuring they possessed the required expertise. Lastly, we prioritized recruiting individuals with a high degree of familiarity with fairness and its related toolkits in a professional context; this was done using custom screening directly in the questionnaire.

Based on (most of) the criteria mentioned above, Prolific identified a pool of 3046 potential participants. We determined the minimum sample size by conducting \emph{a priori} power analysis using G*Power~\cite{faul_2009_GPower}. With an effect size of 15\%, a significance level of 5\%, and a power of 95\%, we calculated the smallest required sample size for seven predictors to be 153 participants. To mitigate potential dropout between the pre-screening and the main survey, we initially collected 201 responses and subsequently received 181 responses for the main survey. The time gap between the two surveys facilitated thorough participant screening. The larger initial pool ensured us a sufficient sample size for analysis, anticipating that some participants might not proceed to the main survey.

Our final sample comprised 181 participants, exhibiting a demographic distribution of 80\% men, 19\% women, and 1\% non-binary individuals. The participants came from 15 diverse countries, with the largest contingents originating from the United States (39\%) and the United Kingdom (24\%). Other nations represented included Australia, Belgium, Canada, Finland, France, Germany, Ireland, Italy, the Netherlands, Portugal, Singapore, Spain, and Sweden, ensuring a rich diversity of perspectives in our study. Our survey encompassed a diverse array of work positions among the 181 respondents. The most prevalent roles included Software Developer/Programmer (25\%), Project Manager (15\%), Data Analyst/Data Engineer/Data Scientist (14\%), and Software Engineer (13\%). 
The respondents also reflected a broad spectrum of ages and experience levels. The majority (52\%) fell within the 30 to 44 years age bracket, followed by 25\% in the 18 to 29 range, and 20\% between 45 and 59 years old. A mere 3\% were older than 69 years. Regarding experience in the software industry, 30\% had over 10 years of experience, 28\% had 1-2 years, and 24\% had 4-6 years. A smaller sample reported 7-9 years of experience (16\%), while only 1\% had less than 1 year of experience in industry.

\subsection{Data Collection}

To facilitate data collection, we developed two questionnaires: the first, designated as the “Pre-Screening questionnaire,” aimed to identify ideal participants—by mean of custom screening—from those already filtered via the Prolific platform. The second, named the “Main questionnaire,” was designed to measure \textsc{UTAUT2} constructs in the participants selected from the pre-screening survey.
Both surveys were developed adhering to the established guidelines by Kitchenham and Pfleeger~\cite{kitchenham2008_PersonalOpinionSurveys} and Andrews et al.~\cite{andrews2007_survey_guidelines}, which are highly regarded in software engineering research. Additionally, we followed the SIGSOFT Empirical Standard for Questionnaire Surveys~\cite{ralph_2020_empirical_standards}. The questionnaires were fully anonymized, featuring an introductory description that provided key details to aid participants in comprehending the tasks. We incorporated a closing question for feedback and attention check questions to ensure participant reliability.


Before administering the surveys, we conducted iterative pilot tests with dual objectives: (1) assessing quality and clarity and (2) estimating completion time. Initially, we orchestrated three pilots involving 10 researchers from our network. After each round, we refined the surveys to address feedback and address any typographical errors. Subsequently, we conducted separate pilots for each questionnaire using Prolific: five participants completed the pre-screening questionnaire, and another five completed the main one. The pilots for both questionnaires happened between 30 August and 01 September 2024.

The pre-screening questionnaire gathered demographic information, assessed participant reliability, and evaluated programming skills and experience. Designed to take 6 minutes to complete, we successfully collected 200 responses within two days starting from 30 August 2024. The main questionnaire, which measured the \textsc{UTAUT2} constructs, was estimated to require 5 minutes for completion, and it took six days to collect 181 responses from 01 September 2024.

\smallskip
\textbf{Ethical Considerations.}
We thoughtfully designed and executed our work, giving careful consideration to participants' privacy and addressing potential ethical concerns inherent in survey studies~\cite{hall2001ethical}.
Our survey design ensured complete anonymity of all responses; consequently, we refrained from collecting participants' names or email addresses. We avoided soliciting any sensitive business information and explicitly guaranteed that the collected data would be utilized solely to address our research objectives. All participants were over 18 years old, provided informed consent prior to participation, and were allowed to withdraw at any time. Moreover, we transparently informed participants that their responses would eventually be published and permanently stored in the online appendix of this paper~\cite{appendix}.

Nevertheless, we acknowledge that gathering insights on critical aspects—such as fairness—from potential employees of organizations that could produce discriminatory ML-based products may still present moral concerns. However, we recognize that industry practitioners have been involved in evaluating fairness and ethics in previous work~\cite{toolkit_survey, organizational_responsible_ai}. Furthermore, given that the scope of the survey was clearly presented in the introduction, we are confident that all participants who responded were genuinely motivated to pursue the cause of providing non-discriminatory solutions.

\subsection{Data Analysis}
As previously explained, data collection was conducted through a survey study. All constructs in the theoretical model described in Section \ref{sec_theory} were measured at the individual level using items validated in the literature, ensuring the reliability of our measurement process. We detail the items used and the data analysis process in the following sections, with a comprehensive overview of all items and references provided in our online appendix \cite{appendix}.

\subsubsection{Data Gathering Instruments}
Questionnaire items assessing the \textsc{UTAUT2} constructs were adapted from the original authors~\cite{utaut2}. The dependent variable, \textit{Use Behavior} (UB), was measured using a single-item frequency scale, while the seven predictors were evaluated on a 7-point Likert scale. These predictors encompassed \textit{Performance Expectancy} (PE, 5 items), \textit{Effort Expectancy} (EE, 6 items), \textit{Social Influence} (SI, 5 items), \textit{Hedonic Motivation} (HM, 3 items), \textit{Facilitating Conditions} (FC, 4 items), \textit{Habit} (HB, 4 items), and \textit{Price Value} (PV, 3 items), as well as \textit{Behavioral Intention} (BI, 3 items). We also asked whether the use of fairness toolkits was mandated by the participants' companies, recognizing this as a potential influencer of use behavior. Additionally, we collected demographic data such as age, gender, role, and years of experience in the software industry to contextualize our sample relative to other surveys.

\subsubsection{Analysis Process}
We initiated our analysis by conducting a thorough preliminary examination of the data to ensure its quality. While PLS-SEM offers considerable flexibility, we nonetheless checked for missing data, unusual response patterns, outliers, and data distribution issues.
Upon validating the dataset, we imported it into SmartPLS, a tool designed specifically for PLS-SEM analysis~\cite{SmartPLS4}. Given the intricate nature of the PLS-SEM process, we direct readers to Hair et al.\cite{hair_2014_PLS} and Russo and Stol\cite{russo2021_pls_SLR} for more detailed explanations.

We began by developing the measurement model (or outer model), which links each theoretical construct to its associated indicators. Each construct in our theoretical framework, i.e., the \textsc{UTAUT2} values, was treated as a latent variable, representing an unobservable concept. Indicators derived from participants' responses were then carefully assigned to their respective constructs. Subsequently, we constructed the structural model (or inner model), which delineates the relationships between these constructs, guided by our hypotheses.

After constructing both the measurement and structural models within SmartPLS and executing the PLS-SEM algorithm, we evaluated both. Given that all our indicators exhibited a reflective relationship with the constructs, we assessed the criteria indicated by Hair et al. \cite{hair_2014_PLS}.

To support replicability and maintain transparency, all materials utilized during the analysis are provided in the comprehensive replication package accompanying this paper \cite{appendix}.
\section{Analysis of the Results}

The following section presents the results from the PLS-SEM analysis. This analysis aims to uncover the causal relationships and underlying patterns within the data, offering a detailed evaluation of the hypothesized model. Through this approach, we gain insights into the interactions between the constructs and validate the theoretical framework proposed.

Before proceeding with the main PLS-SEM analysis, we conducted a preliminary data examination. Notably, there were only a few instances of missing values, likely due to the high quality of the questionnaire and the reliability of our sample, ensured by the approval rate filter. These missing values did not pose any significant issues, as SmartPLS is equipped to manage them automatically. Additionally, we reviewed the data for suspicious response patterns and found none. It is also important to highlight that all participants successfully passed the attention check questions.

\subsection{Measurement Model Evaluation}
As a first step in the evaluation of the theoretical model, it is paramount to evaluate the reliability of the constructs of the model~\cite{hair_2014_PLS, russo2021_pls_SLR}. Consequently, we analyze the indicator reliability, internal consistency reliability, convergent validity, and discriminant validity. This section presents the obtained results for each of the steps mentioned above.

\smallskip
\textbf{Indicator Reliability.}
As outlined by Hair et al.~\cite{hair_2014_PLS}, the initial step in assessing the measurement model is to evaluate the reliability of the indicators, focusing on their \textit{outer loadings}. High outer loadings indicate that the indicators capture a substantial amount of commonality with the construct.

A commonly accepted guideline is to retain indicators with outer loadings above 0.708, while indicators with loadings below 0.40 are generally removed. For those with values between 0.40 and 0.70, removal is considered if it enhances internal consistency reliability or convergent validity.

For brevity, the outer loadings for all indicators are reported in the online appendix \cite{appendix}. Two indicators, EE4 and HB2, had outer loadings below 0.70, but since no indicator had a loading lower than 0.40, all were retained for further analysis.

\smallskip
\textbf{Internal Consistency Reliability.}
The second step involved evaluating \textit{internal consistency reliability} to confirm that the indicators reliably measure their respective constructs. For this assessment, we used three key measures: \textit{Cronbach's alpha}, \textit{composite reliability} ($\rho_c$), and the \textit{reliability coefficient} ($\rho_A$).

The results, presented in Table \ref{table_internal_consistency_reliability}, show that all values exceeded the recommended threshold of 0.60, as suggested by Hair et al.~\cite{hair_2014_PLS}, allowing us to pass this step confidently.

\smallskip
\textbf{Convergent Validity.}
Convergent validity refers to the degree to which a measure correlates positively with other measures of the same construct~\cite{hair_2014_PLS}. Since all constructs in our model utilize reflective indicators, we anticipated that the indicators would converge and share a substantial proportion of variance. The most common metric for assessing this is the \textit{average variance extracted} (AVE). An AVE value of 0.50 or higher is considered acceptable, indicating that the construct captures more than half of the variance of its associated indicators. As reported in Table \ref{table_internal_consistency_reliability}, all constructs in our model have AVE values exceeding the 0.50 threshold, confirming strong convergent validity.

\begin{table}
    \centering
    \normalsize
    \caption{Internal Consistency Reliability and AVE values of the UTAUT2 Constructs.}
    \rowcolors{1}{purple!10}{white}
    \begin{tabular}{l c c c c}
    \hline
        \rowcolor{purple}
        \textcolor{white}{\textbf{Constructs}} & 
        \multicolumn{1}{l}{\textcolor{white}{\textbf{Cronbach's alpha}}} & 
        \multicolumn{1}{c}{\textcolor{white}{{$\mathbf{\rho_A}$}}} & 
        \multicolumn{1}{c}{\textcolor{white}{{$\mathbf{\rho_c}$}}} & 
        \multicolumn{1}{c}{\textcolor{white}{\textbf{AVE}}} 
        \\ \hline
        BI & 0.888 & 0.890 & 0.931 & 0.817 \\ 
        EE & 0.868 & 0.887 & 0.899 & 0.598 \\ 
        FC & 0.817 & 0.829 & 0.879 & 0.647 \\ 
        HB & 0.841 & 0.866 & 0.894 & 0.681 \\ 
        HM & 0.920 & 0.937 & 0.950 & 0.863 \\ 
        PE & 0.915 & 0.921 & 0.936 & 0.746 \\ 
        SI & 0.866 & 0.903 & 0.900 & 0.644 \\
        \hline
    \end{tabular}
    \label{table_internal_consistency_reliability}
\end{table}

\smallskip
\textbf{Discriminant Validity.}
The final test focused on assessing the discriminant validity, which evaluates the degree to which a construct is truly distinct from others. Henseler et al.~\cite{henseler2015_HTMT} introduced the \textit{heterotrait-monotrait ratio} (HTMT) as a reliable criterion for this purpose. HTMT values are computed using the PLS-SEM algorithm, and typically, a value above 0.90 indicates insufficient discriminant validity, while values below 0.85 suggest it is adequate. Additionally, using a \textit{bootstrapping} procedure can further verify whether the HTMT values significantly differ from the threshold. Bootstrapping is a nonparametric technique used to test the significance of various PLS-SEM outcomes, such as path coefficients, Cronbach’s alpha, and HTMT values.

Due to space constraints, the HTMT values derived from our PLS-SEM analysis are included in the online appendix of this paper \cite{appendix}. The results demonstrated that all values fell below the 0.85 threshold. We performed a bootstrapping procedure in SmartPLS with \textit{10000} subsamples, using a one-tailed test at a 0.05 significance level, which confirmed that all HTMT values were below the thresholds. These findings indicate that each construct in our model represents a distinct concept, allowing us to proceed with the evaluation of the structural model.


\subsection{Structural Model Evaluation}
After evaluating the measurement model, the next step is to assess the structural model. 

\smallskip
\textbf{Collinearity Analysis.}
The initial step in evaluating the structural model involves examining collinearity between exogenous and endogenous variables, which is essential for accurate path estimation. To detect multicollinearity, we employed the \textit{Variance Inflation Factor} (VIF), a standard metric used in multiple regression analysis. Ideally, a VIF value under 3 indicates no collinearity, while values below 5 are also acceptable.
In our analysis, the majority of VIF values were below 3, with the highest being 2.47. Only two paths (PE $\rightarrow$ BI and EE $\rightarrow$ BI) slightly exceeded the ideal threshold. Based on these findings, we concluded that multicollinearity does not present a significant concern in our model.

\begin{table*}
    \centering
    \normalsize
    \caption{Significance and Relevance of the UTAUT2 constructs. Significant paths are highlighted in bold and underlined.}
    \rowcolors{1}{purple!10}{white}
    \begin{tabular}{l c c c c c c}
    \hline
        \rowcolor{purple}
        \textcolor{white}{\textbf{Hypotheses}} &
        \multicolumn{1}{l}{\textcolor{white}{\textbf{Path Coefficients}}} &
        \multicolumn{1}{l}{\textcolor{white}{\textbf{Bootstrap Mean}}} &
        \multicolumn{1}{l}{\textcolor{white}{\textbf{St. Dev}}} &
        \multicolumn{1}{l}{\textcolor{white}{\textbf{T statistics}}} &
        \multicolumn{1}{l}{\textcolor{white}{\textbf{P values}}} &
        \multicolumn{1}{l}{\textcolor{white}{\textbf{Significance}}}
        \\ \hline
        \underline{\textbf{BI $\rightarrow$ UB}} & 0.161 & 0.159 & 0.071 & 2.258 & 0.024 & * \\ 
        EE $\rightarrow$ BI & 0.056 & 0.057 & 0.090 & 0.627 & 0.531 & ~ \\ 
        FC $\rightarrow$ BI & 0.090 & 0.098 & 0.077 & 1.169 & 0.242 & ~ \\ 
        FC $\rightarrow$ UB & 0.091 & 0.092 & 0.054 & 1.688 & 0.091 & ~ \\ 
        \underline{\textbf{HB $\rightarrow$ BI}} & 0.323 & 0.319 & 0.091 & 3.531 & 0.000 & **\\ 
        \underline{\textbf{HB $\rightarrow$ UB}}& 0.543 & 0.455 & 0.075 & 6.014 & 0.000 & **\\ 
        HM $\rightarrow$ BI & -0.067 & -0.064 & 0.078 & 0.852 & 0.94 & ~\\ 
        \underline{\textbf{PE $\rightarrow$ BI}} & 0.465 & 0.465 & 0.086 & 5.404 & 0.000 & **\\ 
        SI $\rightarrow$ BI & 0.011 & 0.010 & 0.065 & 0.174 & 0.862 & ~\\  
        \hline
        \rowcolor{white}
        \multicolumn{7}{l}{
        $ 
        {\scriptstyle **: p < 0.001 ;\:\:\: *: p < 0.05}
        $
    }\\
    \end{tabular}
    \label{table_significance_relevance_UTAUT2}
\end{table*}

\smallskip
\textbf{Significance and Relevance of the Relationships.}
In the second phase of our analysis, we focused on evaluating the significance and relevance of the relationships within the structural model. To test for significance, we employed the bootstrapping method, using 10,000 sub-samples, as recommended by Hair et al.~\cite{hair_2014_PLS}. We analyzed T-values, p-values, and bootstrap confidence intervals. The results, summarized in Table \ref{table_significance_relevance_UTAUT2}, indicate that Habit significantly influences both Use Behavior and Behavioral Intention. Furthermore, Behavioral Intention is strongly associated with Use Behavior, while Performance Expectancy demonstrates a significant connection to Behavioral Intention.
To assess the relevance of these significant relationships, we examined the standardized path coefficients, which are also detailed in Table \ref{table_significance_relevance_UTAUT2}. Performance Expectancy emerged as the most influential factor affecting the intention to utilize fairness toolkits, closely followed by Habit. In terms of actual usage behavior among software practitioners, Habit was identified as the most significant factor, with Behavioral Intention ranking second. In addition, further analysis revealed that Performance Expectancy also has an indirect relationship with the use behavior.

\smallskip
\textbf{Explanatory Power.}
In the third phase of our analysis, we aimed to evaluate the model’s explanatory capability, specifically how well it fits the data by quantifying the strength of the relationships within the model, as described by Hair et al.\cite{hair_2014_PLS} and Russo et al.\cite{russo2021_pls_SLR}. This is typically assessed using the coefficient of determination (R\textsuperscript{2}), which ranges from 0 to 1; higher values indicate stronger explanatory power. Although no universal standards exist for R\textsuperscript{2}, values as low as 0.10 may be considered acceptable in certain contexts, with 0.19 often regarded as a more appropriate benchmark~\cite{chin1998_R_Square_value, raithel2012_explanatory_power_PLS_SEM, hair_2014_PLS}.

In our analysis, we found R\textsuperscript{2} values of 0.630 for Behavioral Intention and 0.407 for Use Behavior. This indicates that our model successfully explains 63\% of the variance in the intention to use large language models (LLMs) and 40\% of the variance in actual usage. Furthermore, since R\textsuperscript{2} values are below 0.90, we can confidently exclude the overfitting concern.

After evaluating the coefficient of determination, we further quantified the strength of the relationships using the F\textsuperscript{2} effect size. This metric assesses the potential change in R\textsuperscript{2} if a specific construct were omitted from the model, offering insights into the individual contributions of each construct to the dependent variables.

Focusing on the intention to use fairness toolkits, we found that Performance Expectancy had the largest effect size (0.192), followed by Habit (0.096). When considering actual usage, Habit was the most influential factor (0.150), with Behavioral Intention showing a smaller effect size (0.021). These results highlight the significant influence of Performance Expectancy on the intention to adopt fairness toolkits and underscore the critical role of habitual usage in predicting actual usage.

\smallskip
\textbf{Predictive Power.}
To evaluate the model's practical utility for managerial decision-making, we assessed whether the results derived from our PLS-SEM algorithm are generalizable beyond the specific dataset used in the estimation. This was done using the PLS\textsubscript{predict} procedure~\cite{shmueli2016_PLS_SEM_Predict}, which divides the dataset into training and holdout samples. The key metrics in this analysis were Stone-Geisser's \textit{Q\textsuperscript{2} statistic}, along with the \textit{mean absolute error} (MAE) and the \textit{root mean square error} (RMSE). These values were compared to a benchmark, with Shmueli et al.~\cite{shmueli2016_PLS_SEM_Predict, shmueli2019_PLS_SEM_Predict} recommending a linear regression model (LM) as the benchmark for comparison. A positive Q\textsuperscript{2} value indicates that the model's prediction error is lower than that of the benchmark, while smaller MAE and RMSE values suggest that the model has superior predictive accuracy.

The results, detailed in our online appendix \cite{appendix} due to space constraints, show that all variables outperform the benchmark, indicating that the PLS-SEM model demonstrates strong predictive capabilities.


\stesummarybox{\faBarChart \hspace{0.05cm} Summary of the Results.}{The evaluation of the measurement and structural model confirmed the robustness of the data collection process and allowed us to answer our research question by identifying three key factors driving the adoption of fairness toolkits: Performance Expectancy, Habit, and Behavioral Intention.}

\begin{table*}
    \centering
    \caption{UTAUT2—Summary of Findings and Implications. Significant paths are highlighted in bold and underlined.}
    \rowcolors{1}{purple!10}{white}
    \begin{tabular}{l p{0.4\linewidth} p{0.4\linewidth}}
    \hline
        \rowcolor{purple} 
        \textcolor{white}{\textbf{Hypothesis}} & \textcolor{white}{\textbf{Findings}} & \textcolor{white}{\textbf{Implications}} \\ 
        \hline
        \underline{\textbf{H1: PE $\rightarrow$ BI}} &  This relationship is the strongest of the model regarding the Behavioral Intention, with a path coefficient of 0.465 and an effect size of 0.192. Moreover, this relationship also causes a significant indirect relationship between PE and UB.& Software practitioners' intentions to adopt fairness toolkits are heavily influenced by their expectations of the technology's performance, i.e., how able these instruments are to measure or mitigate bias.
        \\ 
        
        H2: EE $\rightarrow$ BI & The relationship is not significant. & The perceived effort in learning how to apply fairness toolkits to their jobs does not instigate the intention to adopt them.
        \\ 
        
        H3: SI $\rightarrow$ BI & The relationship is not significant. & Practitioners' opinion on the use of fairness toolkits alone does not instigate the intention to adopt the tool.
        \\ 
        
        H4: HM $\rightarrow$ BI & The relationship is not significant. & Practitioners do not intend to adopt fairness toolkits on the basis of the fun and joy their use causes.
        \\ 
        
        H5a: FC $\rightarrow$ BI & The relationship is not significant. & Organizational support and supporting resources do not significantly influence practitioners’ intention to adopt fairness toolkits in their working context.
        \\
        
        H5b: FC $\rightarrow$ UB & The relationship is not significant. & Organizational support and supporting resources do not significantly influence practitioners’ actual adoption of fairness toolkits in their working context.
        \\
        
        \underline{\textbf{H6a: HB $\rightarrow$ BI}} & This relationship is the second most significant regarding Behavioral Intention, with a path coefficient of 0.323 and an effect size of 0.096. & Practitioners that regularly utilize fairness toolkits in their work strengthen their intention to further rely on these tools.\\
        
       \underline{\textbf{H6b: HB $\rightarrow$ UB}} & This relationship is the most significant regarding Use Behavior, with a path coefficient of 0.543 and an effect size of 0.150. A mediation analysis revealed that the effect is direct and not mediated by the relationship with Behavioral Intention. & The habitual use of fairness toolkits can lead to a higher adoption rate.\\

        \underline{\textbf{H7: BI $\rightarrow$ UB}} & This relationship is the second most significant regarding Use Behavior: path coefficient of 0.161 and effect size of 0.021. & As expected, the intention to adopt fairness toolkits results in an actual adoption of the technology.\\
        \hline
    \end{tabular}
    \label{table_discussion_UTAUT2}
\end{table*}

\section{Discussion and Implications}

This study aimed to explore individual factors influencing software practitioners' intention to adopt, as well as their actual adoption, of fairness toolkits using the Unified Theory of Acceptance and Use of Technology (UTAUT) framework. Our findings indicate that three key constructs from the UTAUT2~\cite{utaut2} model—Performance Expectancy, Habit, and Behavioral Intention—exert a statistically significant influence on the dependent variables. Conversely, the other constructs did not show significant effects on the dependent variables. The remainder of this section will explore and elaborate on all the constructs, offering insights and implications that could be valuable for further research and practice.

\subsection{Discussions}
\textit{Performance Expectancy}, or the perceived utility of fairness toolkits, emerges as the cornerstone of adoption. As fairness becomes an increasingly critical non-functional requirement in modern software engineering \cite{brun2018software,mehrabi2021survey}, practitioners are drawn to tools that effectively mitigate bias in ML systems. Our results reveal that this factor has the most significant influence on the intention to adopt fairness toolkits while also maintaining a strong and significant indirect relationship with the actual use behavior of practitioners. These results align with established technology acceptance models \cite{utaut} and recent research in the field \cite{oneto2020mlfairnessbook,figueroa2022ethicalperformanceexpectancy}.

The consideration of fairness in daily workflows is largely driven by \textit{Habit}. The study identifies this as the second most influential factor on \textit{behavioral intention} and the primary determinant of \textit{actual use behavior}.\footnote{Notably, mediation analysis shows that Habit's influence on use behavior is direct and not mediated by \textit{behavioral intention}.} This finding underscores the importance of seamless integration and initial exposure in fostering sustained use of fairness toolkits, as habitual use becomes an essential part of practitioners' routines.

The intentional adoption of fairness toolkits is reflected in the strong relationship between \textit{Behavioral Intention} and \textit{actual use behavior}. This connection, consistent with technology acceptance literature \cite{utaut,utaut2}, emphasizes that cultivating positive intentions through training, awareness programs, and demonstrating benefits can effectively promote adoption \cite{organizational_responsible_ai}.

Notably, the absence of a significant relationship between both \textit{Facilitating Conditions} and \textit{Social Influence} with the intention to adopt and actual use behavior of fairness toolkits suggests that practitioners do not consider organizational support, resource availability, or peer influence as critical factors in their decision to implement these technologies \cite{utaut,thompson1991personal}. This finding challenges conventional assumptions about technology adoption in organizational settings, where such factors are typically seen as essential. Instead, the strong influence of \emph{Performance Expectancy} and \emph{Habits} on the dependent variables indicates that software practitioners are primarily motivated by their perception of the toolkits' effectiveness in mitigating bias and their existing work routines. This suggests a self-driven approach to adoption, where practitioners integrate fairness toolkits based on the anticipated positive impact on their tasks rather than relying on organizational facilitation or social endorsement. Practitioners tend to prioritize the tool's utility and performance, further reducing the impact of social factors on their intention to adopt the technology.

Further supporting this interpretation is the non-significant relationship between \textit{Effort Expectancy} and the intent to adopt these toolkits. This finding suggests that software practitioners, given their technical expertise and familiarity with complex tools, prioritize the functionality and performance benefits of fairness toolkits over considerations of ease of use. It appears that the potential for bias mitigation outweighs concerns about the effort required to implement these technologies\cite{toolkit_survey,toolkit_landscape}. Another explanation could lie in the tools themselves, which may offer such a simple interaction that it becomes effortless for professionals to use. Rather than deterring adoption, this complexity seems to be accepted as an inherent aspect of working with cutting-edge AI ethics solutions. 

Finally, the study found that \textit{Hedonic Motivation} does not significantly influence the \textit{Behavioral Intention} to adopt fairness toolkits. This finding suggests that the potential enjoyment or pleasure derived from using these tools does not play a substantial role in engineers' adoption decisions \cite{utaut}. The non-significance of hedonic motivation in this context is particularly noteworthy. It implies that engineers approach fairness toolkits primarily as utilitarian instruments rather than sources of enjoyment or satisfaction. This perspective aligns with the professional nature of software engineering and the ethical implications of fairness in AI systems. Engineers appear to be driven more by the practical outcomes and ethical considerations of using fairness toolkits than by any intrinsic enjoyment derived from the tools themselves.

Our findings may indicate a high level of professional autonomy and ethical responsibility among software engineers working on AI systems. The emphasis on individual assessment and practical utility in adopting fairness toolkits suggests a workforce that is critically engaged with the ethical implications of their work and committed to addressing these issues through technical means.

\subsection{Implications}
This study contributes to understanding the reasons behind software practitioners' adoption of fairness toolkits. Our results have actionable implications for organizations, toolkit vendors, and researchers.

\smallskip
\textbf{Organizations}. For organizations that aim to spread the usage of fairness toolkits, these insights suggest that efforts to promote adoption should focus on demonstrating their concrete benefits and effectiveness in addressing bias issues. \faHandORight \hspace{0.01cm} \textit{Educational initiatives and awareness campaigns, such as workshops or tutorials}, might be effective if they emphasize the technical merits and the tangible benefits of fairness toolkits rather than relying on social proof, external supports, or attempts to make the tools more enjoyable to use. Moreover, organizations and managers should make an effort to \faHandORight \hspace{0.01cm} \textit{integrate fairness toolkits usage into daily workflows to help employees develop a habit}. This can be done by facilitating access and integration of these tools in daily working activities.

\smallskip
\textbf{Toolkits Vendors}. For toolkits vendors, our work may be of inspiration to understand possible design solutions to enhance the adoption of fairness toolkits. To demonstrate toolkits' high performances, vendors should aim to \faHandORight \hspace{0.01cm} \textit{provide practical examples and real-world cases} in which their solutions helped mitigate biases and achieve fair ML models, rather than relying on theoretical proofs. In addition, to make practitioners develop a habit of the use of fairness toolkits, vendors should \faHandORight \hspace{0.01cm} \textit{facilitate practitioners in integrating such solutions in their daily activities} through efficient APIs or libraries. Despite the efforts vendors can make to promote toolkit adoption, our findings indicate that practitioners perceive these tools as both useful and effective. This suggests that the investment in supporting fair ML development is paying off.

\smallskip
\textbf{Researchers}. Finally, researchers should leverage our findings to perform further investigations on fairness toolkits. On the one hand, \faHandORight \hspace{0.01cm} \textit{empirical studies demonstrating these solutions' performances and abilities in mitigating bias} could further increase practitioners' intention to adopt them. On the other hand, exploring \faHandORight \hspace{0.01cm} \textit{novel ways to integrate and automate fairness toolkits' integration in existing workflows}, such as CI/CD pipelines, could tempt software practitioners to use them and consequently develop a habit.

\section{Threats To Validity}
Our study primarily focused on quantitative analysis supported by statistical methods. In discussing threats to validity, we followed the framework outlined by Wohlin et al.~\cite{wohlin2012_experimentation}.

Regarding the \textbf{conclusion Validity}—i.e., threats about the ability to draw accurate conclusions about the relationships between independent and dependent variables~\cite{wohlin2012_experimentation}—the primary threats in this category stem from the statistical tests used for analysis. To address this, we relied on PLS-SEM, which is known for its robustness in various contexts. We closely followed the procedures outlined by Hair et al.\cite{hair_2014_PLS} in their detailed work on PLS-SEM methodology. Moreover, we employed SmartPLS, a widely used software cited in over 1,000 peer-reviewed studies\cite{hair_2014_PLS}.

Concerning \textbf{internal Validity}—i.e., the risk that external factors may have influenced the dependent variable, leading to inaccurate conclusions~\cite{wohlin2012_experimentation}—we grounded our study in well-established theories to avoid it. Indeed, we utilized the Unified Theory of Acceptance and Use of Technology, which is specifically suited for investigating phenomena of this nature~\cite{utaut2}. We also applied filters to our participant sample to ensure it accurately represented the target population while maintaining sufficient diversity.

Moving on \textbf{construct Validity}—that concerns about the accuracy of the measurements and tools used to represent the study variables~\cite{wohlin2012_experimentation}—all variables were assessed using validated instruments~\cite{utaut2}. The questionnaires were designed in line with the latest field guidelines, and we employed strategies such as question randomization and attention-checks to improve the reliability of the results~\cite{kitchenham2008_PersonalOpinionSurveys, ralph_2020_empirical_standards, danilova2021_developers_questions}.

Last, we tried to address \textbf{external Validity}—that is about the generalizability of the findings to a broader population~\cite{wohlin2012_experimentation} filtering the Prolific population to select participants with characteristics aligned to our study’s objectives. Additionally, we gathered sufficient data in line with G*Power recommendations~\cite{faul_2009_GPower}. While the majority of participants were from Europe, which reflects Prolific's user distribution, we acknowledge this limitation. Nonetheless, we believe the results offer valuable insights.
\section{Conclusion}

This study investigated the adoption of fairness toolkits among software practitioners using the Unified Theory of Acceptance and Use of Technology (UTAUT2) framework~\cite{utaut2}. We surveyed experts and analyzed the data using Partial Least Squares Structural Equation Modeling (PLS-SEM)~\cite{hair_2014_PLS}.

Our findings reveal that \textit{Habit} and \textit{Performance Expectancy} significantly influence the intention to adopt fairness toolkits, aligning with previous research~\cite{oneto2020mlfairnessbook,figueroa2022ethicalperformanceexpectancy}. Moreover, \textit{Habit} emerged as the primary driver for the actual use of these toolkits, alongside practitioners' intention to use them.

These results have important implications. Organizations promoting fairness toolkit adoption should focus on demonstrating their concrete benefits and effectiveness in addressing bias issues. Additionally, our findings suggest that software practitioners primarily approach bias mitigation from a technical perspective, indicating a continued need for research into algorithmic solutions for ML fairness.

Future research should explore the impact of additional factors, such as cultural values, on the adoption of these technologies. We also recommend longitudinal studies to understand how these results may evolve as the technology matures and awareness of AI ethics grows within the software development community.

\section*{Data Availability}
The data that support the findings of this study are openly available in our online appendix \cite{appendix}.

\section*{Acknowledgment}
We acknowledge the use of ChatGPT-4 to ensure linguistic accuracy and enhance the readability of this article. This work has been partially supported by the European Union - NextGenerationEU through the Italian Ministry of University and Research, Projects PRIN 2022 PNRR "FRINGE: context-aware FaiRness engineerING in complex software systEms" (grant n. P2022553SL, CUP: D53D23017340001). The opinions presented in this article solely belong to the author(s) and do not necessarily reflect those of the European Union or The European Research Executive Agency. The European Union and the granting authority cannot be held accountable for these views.

\balance
\bibliographystyle{IEEEtran}
\bibliography{references}

\end{document}